\documentclass{elsart}

 \usepackage{graphicx}
 \usepackage{epsfig}

\usepackage{amssymb}

\begin{document}

\begin{frontmatter}


 \title{Jamming transition in air transportation networks}
 \author{Lucas Lacasa$^{1,2}$, Miguel Cea$^1$ and Massimiliano Zanin$^1$}
\address{$^1$ Innaxis Foundation $\&$ Research Institute\\
Vel\'{a}zquez 157, Madrid, Spain.\\
$^2$Departamento de Matem{\'a}tica Aplicada y Estad{\'i}stica\\
ETSI Aeron{\'a}uticos\\ Universidad Polit{\'e}cnica de Madrid, Spain.\\
}

\begin{abstract}
\noindent In this work we present a model of an air transportation
traffic system from the complex network modelling viewpoint. In
the network, every node corresponds to a given airport, and two
nodes are connected by means of flight routes. Each node is
weighted according to its load capacity, and links are weighted
according to the Euclidean distance that separates each pair of
nodes. Local rules describing the behavior of individual nodes in
terms of the surrounding flow have been also modelled, and a
random network topology has been chosen in a baseline approach.
Numerical simulations describing the diffusion of a given number
of agents (aircraft) in this network show the onset of a jamming
transition that distinguishes an efficient regime with null amount
of airport queues and high diffusivity (free phase) and a regime
where bottlenecks suddenly take place, leading to a poor aircraft
diffusion (congested phase). Fluctuations are maximal around the
congestion threshold, suggesting that the transition is critical.
We then proceed by exploring the robustness of our results in
neutral random topologies by embedding the model in heterogeneous
networks. Specifically, we make use of the European air
transportation network formed by $858$ airports and $11170$ flight
routes connecting them, which we show to be scale-free. The
jamming transition is also observed in this case. These results
and methodologies may introduce relevant decision making
procedures in order to optimize the air transportation traffic.
\end{abstract}


\end{frontmatter}

\section{Introduction}

In the last decades Physics of Complex Systems and Complexity
Science has started to address real world problems. In particular,
much attention has been paid to self-driven particles such as
pedestrian and freeway traffic \cite{ACreview1,ACreview2} or
Internet traffic systems \cite{INT0}. Not that surprisingly,
complex features such as the emergence of phase transitions
\cite{INT0,TF0,TF1,TF2,TF3} or criticality \cite{INT2} take place
in models that characterize these complex collective phenomena.
Thereby, such complex systems focuses seem to be both a realistic
and useful approach when describing the concept of traffic
dynamics \cite{ACreview1}, both in homogeneous and in
heterogeneous media.

The first insights considering traffic dynamics from a cooperative
phenomenon point of view were developed in cellular automata. In
\cite{ACS} and subsequent works, Nagel and Shreckenberg developed
a stochastic discrete model of freeway traffic dynamics which
evidenced a free-congested phase transition quite similar to the
real traffic behaviors. Successive refinements and generalizations
of this model such as \cite{TF0} or \cite{TF1,TF3} have been
performed so far. All these models focus on how the aggregation of
local dynamical rules may generate emergent nonlinear behaviors at
the global level, such as travelling jam waves, for instance.
Pedestrian dynamics has also been addressed from a complex system
point of view. Self-organization effects occurring in pedestrian
crowds which lead to unexpected mutual disturbances of pedestrian
flows, such as panic effects \cite{HELB1}, have been recently
studied.

On the other hand, in recent years much attention has been paid to
another type of traffic dynamics, taking place in heterogeneous
media: the Internet traffic. Similar insights have been
considered: free/congested phase transitions have been observed,
explaining the real behavior of the Internet performance (for
instance, \cite{INT0} and subsequent works). Other complexity
features such as information transfer in the transition
neighborhood \cite{INT1} or self-organized criticality \cite{INT2}
have been also addressed in these systems. Some evident analogies
between highway and Internet systems have even been put forward
\cite{INT3}. The main difference between both hallmarks is that in
the case of Internet traffic models, the topology of the
underlying network of interactions \cite{SFN} is inevitably
coupled to the internal traffic dynamics, in such a way that the
emergent behavior that takes place is related to the interplay
between complex collective dynamics and complex interaction
topologies. As a matter of fact, it is now well known that complex
interaction topologies strongly affect the dynamics (for instance,
percolation and epidemic thresholds have been found to be highly
topology-dependent \cite{TOPO1,TOPO2,TOPO3,TOPO4,TOPO5}), so that
proper modelling of such phenomena should take good care of this
fact.

Here we will handle a system quite similar to the former ones: the
Air Transportation System (ATS). This one is formed by a spatially
extended network that physically covers a wide range of the world,
made up by weighted nodes (airports of different characteristics)
and links (flight routes) through which aircraft flows diffuse. As
far as the latter system shows striking similarities with the
highway system or Internet, in terms of non-linear coupling of
local dynamics, queueing generation and congestion propagation
phenomena, it is straightforward that the ATS merits a complex
system insight. Some recent studies have focused on the air
transportation network topology \cite{sfn2} and its structure
\cite{sfn3,bagler} or on its application to real epidemic
spreading \cite{epi0,epi}.
 Much on the contrary, the air navigation modelling state-of-art focusses on local models and
 uncoupled network models (for instance, \cite{ATM1,ATM2,ATM3,ATM4}),
 rather than global models that take into account the nonlinear coupling
effects. While some traffic dynamics systems in complex topologies
have been recently performed \cite{Hu,zhao,GEN}, specific systems
that address the Air traffic modelling are somehow lacking. In
this paper we present a network based model of the ATS that
simulates the effect of traffic dynamics. In section II we present
the model, in terms of the network definition and the local
dynamical rules. A random network topology is chosen in a baseline
approach, in order to study the effect of local dynamical rules in
the global behavior without any further source of complexity. In
section III we point out the emergence of a jamming transition in
the dynamics of aircraft diffusion, which distinguishes a regime
where the average amount of queues in the network is null
(efficient regime) from a regime where this average value is non
null due to bottleneck generation (inefficient regime). Moreover,
we show that the transition is critical. In section IV we extend
the neutral model by embedding the system in heterogeneous
networks. To this end we generate the (real) European air
transportation network, formed by $858$ European airports and
$11170$ flight routes connecting them. We firstly show that this
network exhibits a scale-free topology, in good agreement with
previous results for the worldwide transportation network
\cite{sfn2} (indeed we show that similar exponents appear). The
simulations suggest that the dynamics in this real network are
qualitatively similar to what is found for the random (neutral)
topology, what means that the neutral model is -at least for
networks of comparable sizes- robust against changes in the
network topology. In section V we provide some conclusions and
depict some further work.

\section{The model}

We will model the ATS as a complex directed network where some
dynamics take place. In the network, each node is an airport, and
two nodes are linked if there exist a flight route between them
(note that between two linked nodes, both directions are defined).
Each node is weighted in order to characterize the airport's
design capacity (the design capacity stands for the maximal number
of aircrafts per time unit that an airport can handle in an ideal
situation). Each link will also be weighted in order to implement
a metric layer in the network: each link weights characterizes the
Euclidean distance between node pairs. Instead of the adjacency
matrix alone, which fully characterizes an undirected graph
\cite{graphs}, this co-weighted directed network will be
characterized by a triple, formed by an adjacency matrix (that
describes the topology structure), a distance matrix (which
describes the geometric structure and the link weights) and a
design capacity vector (which describes the node weights). As a
neutral model, we have chosen a random network topology
\cite{graphs}. Observe at this point that while further relaxation
of this neutrality will be performed in the section IV (where we
run the model in a realistic non Poissonian network), it is
necessary in a baseline approach to run the model in a network
whose topological complexity may not have an effect in the global
dynamics. This random network has $n=100$ nodes where:\\

(i) The node's weights (design capacities) are chosen randomly from a uniform distribution $U[1,1000]$
and are fixed for different network geometries.\\
(ii) The nodes have been linked randomly, with a link probability $p=0.2$
(the mean degree is consequently $<k>=p(n-1)/2\simeq 10$).\\
(iii) The link's weights are integers chosen randomly from a uniform distribution $U[1,10]$,
characterizing the number of time steps that a given flow will need to cover the distance between two given nodes.\\

In addition, the real capacity $RC$ of each node will be updated
each time step as a percentage of its design capacity $DC$,
modelled stochastically in the following way:
\begin{equation}
RC(t)=DC(1-\xi), \label{capacidad_real}
\end{equation}
where $\xi$ is a random variable extracted from a uniform
distribution $U[0,z]$, $z$ characterizing the noise level.
Note that this modelling is quite intuitive: the performance of individual airports deviates from their design values due to a large number of variables (changing weather conditions, runways availability, unexpected failure of software/hardware infrastructure, human related effects, to cite but a few) that can be simulated by a fitted random variable \cite{random}. Therefore, when $z$ is small, we assume that the node performance indicators are quite reliable and that unexpected events are not likely to occur. A large value of $z$ denotes, on the contrary, a node restricted to high fluctuations in its performance indicators.\\
An initial number of aircrafts (aircraft density) is distributed uniformly over the network.
Then these flows diffuse between the nodes along the links. When flows travelling  across different links reach a given node $i$,
the local rules which apply are the following:\\

(i) The input flow $IF$ of the node $i$ is a sum of the corresponding flows coming from $i$'s nearest
neighbors that reach $i$ in that step (note that distance effects are relevant at this point).\\
(ii) There is a balance between the input flow and the queue generation in the following terms: if
the input flow exceeds the real capacity, an amount equal to the real capacity will constitute the output flow (OF)
and the difference will remain as a queue $IF>RC \rightarrow OF=RC$. The output flow is then
distributed in the out-links proportionally to the weights of the corresponding target nodes.\\
(iii) If the input flow is smaller than the node's real capacity the output flow
is then formed by the input flow plus any remaining queue flow $Q$ (until reaching the real capacity),
$IF<RC \rightarrow OF=IF+Q$.\\
(iv) When the queue of any nearest neighbor's node exceeds a given threshold
$Q_{t1}$, the real capacity $RC$ of node $i$ reduces to zero in that step provided that its
own queue does not exceed a given value $Q_{t2}$ (network coupling rule). This rule mimics the
following effect: a flight departure connecting two airports will be delayed if the target airport
is congested ($Q_{t1}$). However if the departure airport is congested too, this aircraft will need
to get out anyways ($Q_{t2}$).\\

Note that the updating is parallel and that as long as there is a
metric layer defined in the network, the nodes updating is not
synchronized.

\section{Phase transition}
As the flow diffuses along the network, the interplay between the
network's topology, the distance desynchronization effects, the
capacities stochastic updating and the coupling local rules lead to
highly nonlinear dynamics. In order to study the global behavior of
the network, we may define $Q_i(t)$ as the evolving queue amount at
node $i$ and $P$ as the percentage of aircrafts that are not stuck
in a node's queue, measured in the steady state. Note that $P$
actually measures the network's efficiency as far as it gives the
flow rate which is diffusing as compared to the flow rate which is
stuck.
\begin{figure}
\centering
\includegraphics[width=0.70\textwidth]{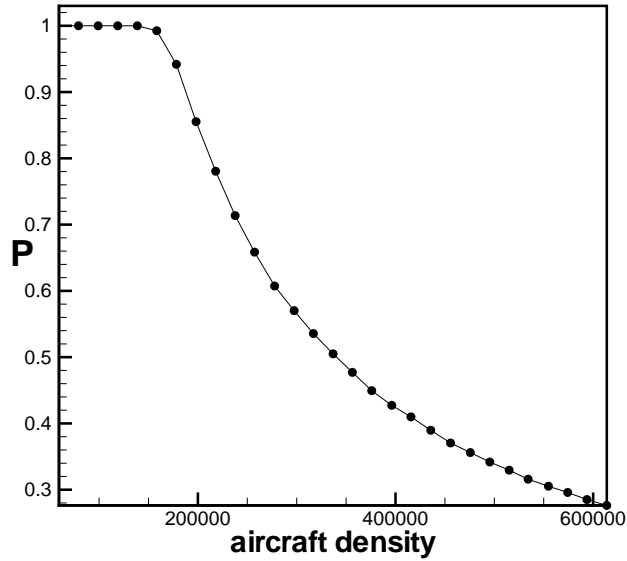}
\caption{Phase diagram relating the percentage of diffusing
aircrafts $P$ as a function of the network aircraft density (each
dot is the average of $500$ network realizations). Two separated
regimes arise, namely: a free (laminar) phase for low densities
which stands for the efficient regime, where every aircraft is
constantly diffusing along the network without any queue constraint
($P=1$), and a congested (turbulent) phase for high densities where
bottlenecks arise (inefficient regime) and the aircraft diffusivity
sharply decreases, giving rise to a jamming transition.}
\label{fig.2}
\end{figure}
We have run several Monte Carlo simulations of the diffusion of a
number of aircraft (the so called aircraft density) all over the
network and have measured the queue generation pattern. In figure
\ref{fig.2} we have represented the steady values of $P$, as a
function of the total aircraft density (note that results have
been averaged over $500$ network realizations). For these concrete
simulations we have set $z=0.1$, $Q_{t1}=4000$, and $Q_{t2}=4000$.
Note that two separated regimes arise, the first where every
aircraft is constantly diffusing (null amount of queues and
consequently $P=1$) and the second where the amount of queues is
non-null and the percentage of diffusing aircrafts decreases,
giving rise to a sort of jamming transition. We will quote the
free phase as the efficient phase, as far as a null amount of
queues denotes an efficient network. On the opposite, the
congested phase will be quoted as the inefficient phase. As far as
$P$ characterizes the rate of aircraft which are diffusing, we can
deduce that there is a threshold that distinguishes free moving
behavior from congestion propagation. Figure \ref{fig.2} is indeed
related to the so-called fundamental diagram in the traffic
dynamics literature \cite{ACreview1}. Such pattern of
free/congested behavior have already been described in both
highway traffic and in computer network traffic models, as
commented in the introduction section. In the air transport
hallmark, this behavior can be interpreted in the following terms:
if the public demand (or eventually the slot adjudication) exceeds
a given threshold in terms of number of demanded flights/slots per
day, the network will decrease its performance in terms of delay
rates increase in a prominent way. This kind of studies may thus
constitute a key indicator of the
adequate balance between public flight demand and capacity supply.\\

\begin{figure}
\includegraphics[width=0.70\textwidth]{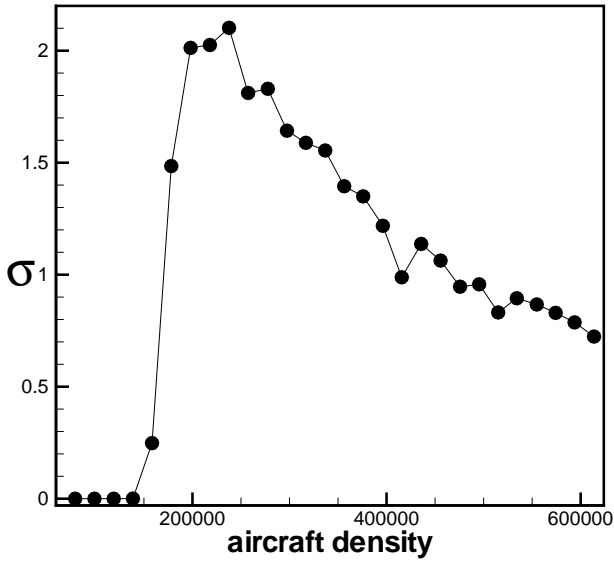}
\caption{Variance of $P$, measuring the system's fluctuations, for
the same simulations as figure \ref{fig.2} (results have been
averaged over $500$ network realizations). Note that $\sigma$
reaches a maximum in a neighborhood of the transition, what is
characteristic of a critical phenomenon (any shift is due to finite
size effects). System's uncertainty becomes maximal in this region.}
\label{var}
\end{figure}

\begin{figure}
\includegraphics[width=0.70\textwidth, angle=-90]{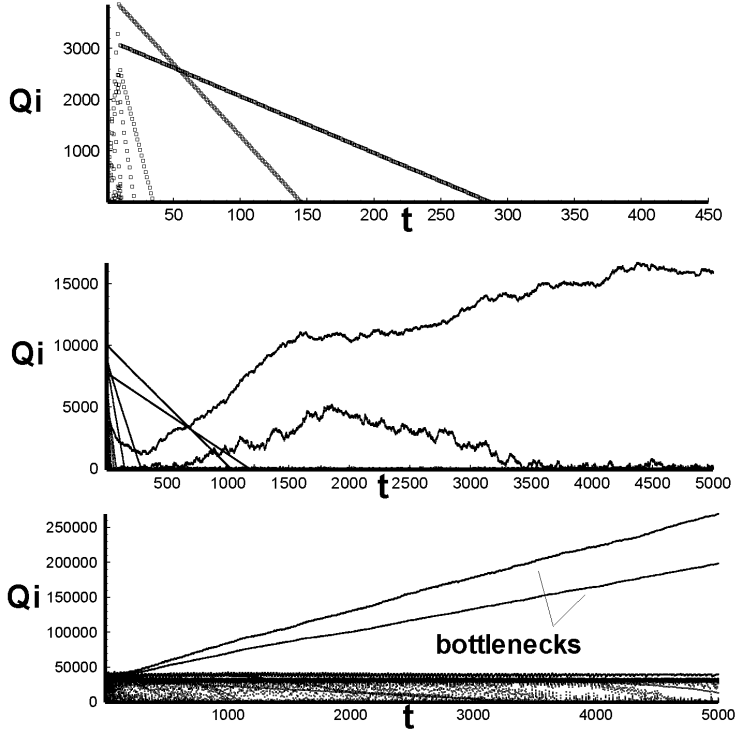}
\caption{Temporal evolution of the node's queues, for three
different initial aircraft densities: (up) for low aircraft
densities, the system quickly evolves into a free phase with no
presence o queues, (middle) for medium aircraft densities, almost
every node's queue converges to a null value except for some nodes
where the queues evolution mimic a brownian-like motion, (down) for
high aircraft densities, the system is quickly congested due to the
birth of several bottlenecks.} \label{fig.3}
\end{figure}

\begin{figure}
\includegraphics[width=0.70\textwidth]{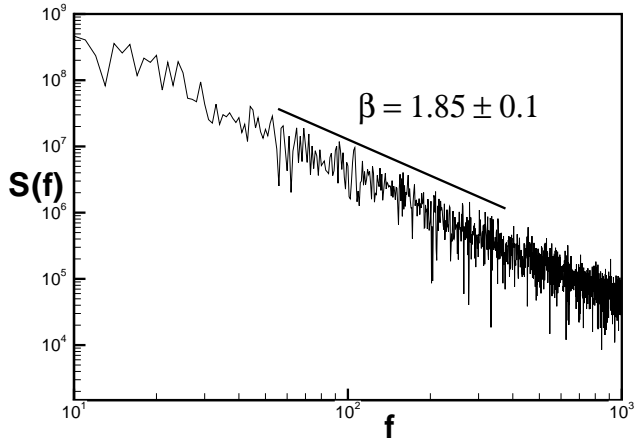}
\caption{Power spectral density $S(f)$ of the temporal evolution
of a node's queue in the critical state. The plot is log-log: the
node's queue displays fractal behavior as long as the power
spectrum is of the shape $S(f)\sim1/f^\beta$, where the best
fitting provides $\beta = 1.85\pm0.1$.} \label{spectrum}
\end{figure}

In order to provide a deeper insight of the dynamical process
taking place in the system, in figure \ref{var} we have plotted
the variance of $P$, $\sigma$, as a function of the aircraft
density, for the same simulations as for figure \ref{fig.2}. This
variable characterizes the fluctuations that the system has, that
is to say, the sensibility of the system regards to small
perturbations, i.e. its uncertainty \cite{INT1}. Note that the
variance reaches a peaked maximum in a neighborhood of the jamming
transition (any shift is due to finite size effects): the system
reaches the maximal degree of unpredictability close to the
transition point. This critical behavior is in turn suggesting
that some information measures -such as the information
entropy- may reach a maximum in this region.\\
We may also shed light into the microscopic dynamical evolution of
the system: in figure \ref{fig.3} we have plotted the amount of
queues in every node as a function of time $Q_i(t)$, for three
different aircraft density conditions. In the upper part of the
figure, the network aircraft density has been set such that the
system belongs to the free phase (according to figure
\ref{fig.2}). Note that $Q_i(t)$ quickly evolves to a null value
(the network is able to absorb the transient queues and the system
quickly evolves into a free moving regime): in this situation the
aircraft flows are constantly diffusing over the network without
any queue constraint. In the bottom part of the figure, the
network is in the congested phase. We can see how after a
transient where many node's queues converge to a null or confined
value, different nodes convert in bottlenecks, such that its
queues increase monotonically in time. In this situation, a high
amount of aircrafts are stuck in the bottlenecks queues, and as a
consequence the flow diffusion is poor. In the middle part of the
figure, the network regime is in a neighborhood of the critical
point (according to figure \ref{fig.2}). We can see how the
majority of the node's queues still converge to a null value,
except for some nodes, whose queues are large and oscillate in an
erratic trend, much in the way of a fractal random walk (fractal
behavior is typical of a critical state). The evolving values of
$Q_i(t)$ are indeed time series. In figure \ref{spectrum} we have
represented a spectral analysis of one of these erratic time
series. We have plotted the values of the Power spectral density
$S(f)$ in a log-log scale: the system clearly evidences
$1/f^\beta$ noise, as long as $S(f)\sim f^{-\beta}$ with $\beta =
1.85\pm0.1$. This feature clearly indicates that the system is
critical, in the sense that it is highly sensible to small
perturbations and is consequently more unpredictable, as seen in figure \ref{var}.\\

\section{Real data: European air transportation network}
\begin{figure}
\centering
\includegraphics[width=0.70\textwidth]{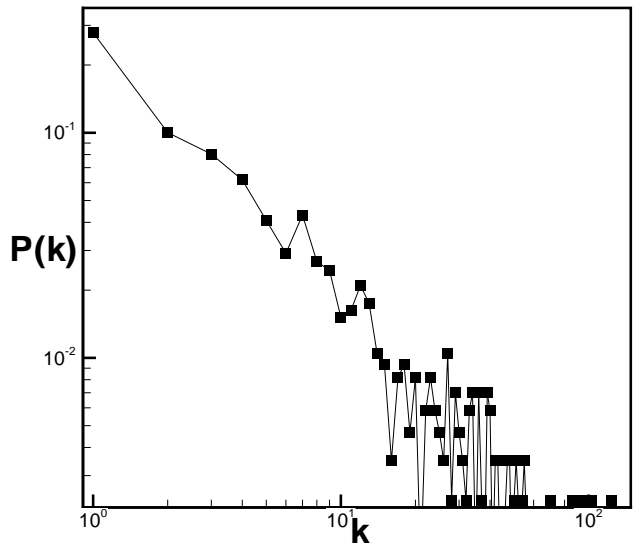}
\caption{Plot in log-log of the degree distribution $P(k)$ of the
European air transportation network, formed by $858$ European
airports connected by $11170$ actual flight routes. The network
shows scale-free behavior $P(k)\sim k^{-\alpha},\ \alpha=1.1\pm0.1
$, in good agreement with what was previously found for the
worldwide air transportation network \cite{sfn2}.} \label{fig.5}
\end{figure}

Real transportation networks topologies differ from random
Poissonian topology as their degree distribution seems to be
heavy-tailed \cite{sfn2,sfn3}. Since dynamics running on different
topologies may eventually give rise to different macroscopic
behaviors, we chose a random topology as the neutral embedding
network. Here, in order to study the preceding model in more
realistic scenarios, we have embedded our model in a real air
transportation network. The nodes of this network represent the
$858$ relevant European airports (in terms of aircraft capacity
and actual traffic). The adjacency matrix encodes real flight
routes while the distance matrix encodes the Euclidean distance
between each pair of connected airports (note that the effect of
the earth's curvature has been taken into account in the distance
calculations). Moreover, the airports weights (design capacities)
have been
estimated with real traffic data.\\
Prior to simulating the traffic dynamics, we have calculated some
topological features of this European network. Concretely, in
figure \ref{fig.5} we have plotted in log-log the degree
distribution $P(k)$ of this network. A power law relation
$P(k)\sim k^{-\alpha}$ holds: the European transportation network
seems to be scale-free, in good agreement with topological
signatures of the worldwide air transportation network \cite{sfn2}
(note that as the network is not too large -$858$ nodes-,
statistics evidence some finite size effects as noisy tails) and
the Indian network \cite{bagler}. The slope $\alpha=1.1\pm0.1$
which is, interestingly, in good agreement with the one previously
found for the worldwide air transportation network \cite{sfn2} but
different from the one for the Indian case \cite{bagler}. This
fact may eventually suggest that similar mechanisms hold for the
growing of worldwide and European networks. Furthermore, note that
in the neutral model, the design capacity of each airport was a
random variable extracted from a uniform distribution. However, in
the real network the nodes weight distribution (that is, the
distribution of design capacities) is no longer uniform but
heavy-tailed, since the capacity of an airport is correlated with
the number of flight routes that it possesses, that is, its degree
(the values have been extracted from traffic data providing the
European airport's
daily activity).\\

\begin{figure}
\centering
\includegraphics[width=0.70\textwidth]{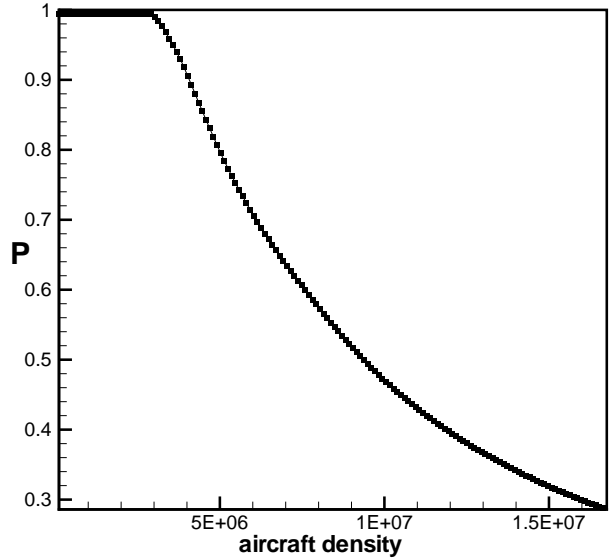}
\caption{Phase diagram relating the percentage of diffusing
aircrafts $P$ as a function of the network aircraft density, as
result of the Monte Carlo simulation of the traffic model embedded
in the European air transportation network (scale-free network of
formed by $858$ nodes). Results are qualitatively similar to those
found within a random network.} \label{fig.6}
\end{figure}

\noindent Then, we have simulated the diffusion of a given number
of aircraft through this network according to our traffic model,
for values of $Q_{t1}=4000$, $Q_{t2}=4000$ and $z=0.1$ (the same
values as for the neutral model). The steady values of the
network's efficiency $P$ versus the total number of aircraft is
plotted in figure \ref{fig.6}. Note that the results are
qualitatively similar as those found for the random (neutral)
topology: above a certain threshold, the aircraft diffusion
sharply decreases due to congestion effects. Quantitative
differences between both scenarios are however not easy to point
out.

\section{Summary}
In this work we have introduced a network based model which
simulates the Air Transportation traffic dynamics. Coupled to the
network definition, local rules modelling the airport's behavior
have been defined in the nodes. As a baseline study, a random
network topology with a metric layer has been chosen. A critical
transition distinguishing a free diffusing aircraft regime from a
congested one has been found numerically. This behavior is robust
against changes in the network topology, since simulations running
in real air transportation networks (which we have found are
scale-free) are qualitatively similar to those obtained in random
(neutral) topologies. Further work should refine this baseline
traffic model, in terms of more realistic rules and scenarios,
systematic analysis of CDM rules, different noise implementations,
and quantitative analysis of delay patterns. Note that the time
restrictions that a schedule introduce in the network's
availability may also have an important effect on the system
dynamics \cite{max}, an issue that will also be investigated in the future.\\
In any case, this kind of methodologies focusing on complex
cooperative behavior may be a starting point for the development
of strategies and decision making procedures in order to optimize
the dynamics of the Air Transportation System, as well as other
network-based transportation systems.

\section{acknowledgments} The authors thank B. Luque, F.J.
Mancebo, S. Cristobal and the Innaxis modelling group for helpful
suggestions, as well as two anonymous referees for their comments.

\end{document}